\documentclass[]{aa}

\def\lesssim{\mathrel{\hbox{\rlap{\hbox{\lower4pt\hbox{$\sim$}}}\hbox{$<$}}}}

\def\gtrsim{\mathrel{\hbox{\rlap{\hbox{\lower4pt\hbox{$\sim$}}}\hbox{$>$}}}}

\def\msun{M$_{\odot}$}

\def\ll_lsun{log$({L/\rm L-_{\odot}})$~}

\def\masa_msun{$M/ \rm M_{\odot}$~}

\def\m_mstar{$M/M_{*}$~}

\usepackage{graphicx}
\usepackage{txfonts}        

\begin{document}
\title{New  evolutionary calculations for  the Born  Again scenario} 

\author{M. M. Miller Bertolami$^1$\thanks{Fellow of CONICET, and IALP
        CONICET/FCAG-UNLP, Argentina}, L. G. Althaus$^1$\thanks{Member
        of the Carrera del Investigador Cient\'{\i}fico y
        Tecnol\'ogico and IALP, CONICET/FCAG-UNLP, Argentina.},
        A. M. Serenelli$^2$, and J. A. Panei $^1$\thanks{Fellow of
        CONICET, and IALP CONICET/FCAG-UNLP, Argentina}}
        \offprints{mmiller@fcaglp.unlp.edu.ar}

\institute{ $^1$ Facultad de Ciencias Astron\'omicas y
Geof\'{\i}sicas, Universidad Nacional de La Plata, Paseo del Bosque
S/N, (B1900FWA) La Plata, Argentina\\ $^2$ Institute for Advanced
Study, School of Natural Sciences, Einstein Drive, Princeton, NJ, 08540, USA\\
\email{althaus,panei,mmiller@fcaglp.unlp.edu.ar,aldos@ias.edu}}

\date{Received; accepted} 

\abstract{We present evolutionary calculations aimed at describing the
born-again scenario for post-AGB remnant stars of 0.5842 and 0.5885
\msun. Results are based on a detailed treatment of the physical
processes responsible for the chemical abundance changes. We
considered two theories of convection: the standard mixing length
theory (MLT) and the double-diffusive GNA convection developed by
Grossman et al. The latter accounts for the effect of the chemical
gradient ($\nabla\mu$) in the mixing processes and in the transport of
energy.  We also explore the dependence of the born-again evolution on
some physical hypothesis, such as the effect of the existence of
non-zero chemical gradients, the prescription for the velocity of the
convective elements and the size of the overshooting zones. Attention
is given to the behavior of the born-again times and to the chemical
evolution during the ingestion of protons.   We find that in our
calculations born again times are dependent on time
resolution. In particular when the minimum allowed time step is
below $5 \times 10^{-5}$ yr  we obtain, with the standard mixing
length theory, born again times of 5-10 yr. This is true without
altering the prescription for the efficiency of convective mixing
during the proton ingestion. On the other hand we find that the
inclusion of the chemical gradients in the calculation of
the mixing velocity tend to increase the born again times by about a
factor of two. In addition we find that proton ingestion can be
seriously altered if the occurrence of overshooting is modified by the
$\nabla\mu$-barrier at the H-He interface, strongly altering born again times.
}  
\authorrunning{Miller Bertolami et al.}
\titlerunning{New evolutionary calculations for the born again
scenario} \maketitle


\section{Introduction}
The surface chemical composition that characterizes post-asymptotic
giang branch (AGB) stars is diverse and poses a real challenge to the
stellar evolution theory.  In particular, about 20\% of these objects
exhibit hydrogen(H)-deficient surface compositions.  The most widely
accepted mechanism for the formation of H-deficient stars remains the born
again scenario, which develops as a result of a last helium thermal
pulse that occurs after the star has left the thermally pulsing
AGB. If this pulse happens when H-burning has almost ceased on the
early white-dwarf cooling branch, it is classified as a very late
thermal pulse (VLTP). In a VLTP, the outward-growing convection zone
powered by helium burning reaches the base of the H-rich envelope.  As
a result, most of H is transported downwards to the hot helium burning
region, where it is completely burned.

Observational examples of stars that are believed to have experienced
a born-again episode are the oxygen-rich, H-deficient PG-1159
stars and the Wolf Rayet type central stars of planetary nebulae
having spectral type [WC] (Dreizler \& Heber 1998 and Werner
2001). In particular, the high surface oxygen abundance detected in
these stars has been successfully explained in terms of overshoot
episodes below the helium-flash convection zone during the thermally
pulsing AGB phase (Herwig et al. 1999, see also Althaus et al
2005).

Both theoretical and observational evidence (Dreizler \& Werner 1996,
Unglaub \& Bues 2000 and Althaus et al. 2005) suggest an evolutionary
connection between most of PG1159 stars and the helium-rich DO stars,
the hot and immediate progenitors of the majority of DB white dwarfs.
Thus, the born again process turns out to be a key phenomenon in
explaining the existence of the majority of DB white dwarfs.  In
addition, the identification of the Sakurai's object (V4334 Sgr) as a
star just emerging from a born-again episode (Duerbeck \& Benetti
1996) has led to a renewed interest in assessing the evolutionary
stages corresponding to the VLTP and the ensuing born-again. V4334 Sgr
has been observed to suffer from an extremely fast evolution. Indeed,
it has evolved from the pre-white dwarf stage into a giant star in
only about a few years (see Asplund et al. 1999, Duerbeck et
al. 2000).  Also recent observations (Hajduk et al. 2005) seem to
indicate that Sakurai's object is quickly reheating. Unfortunately,
very few detailed numerical simulations through the born-again stage
regime exist in the literature. Evolutionary calculations that
incorporate appropriate time-dependent mixing procedures were carried
out initially by Iben \& MacDonald (1995), who found short born-again
time scales of about 17 yr.  By contrast, Herwig et al.(1999) and
Lawlor \& MacDonald (2003) derived too large born-again timescales,
typically 350 yr, to be consistent with observations.  This has
prompted Herwig (2001) to suggest that, in order to reproduce the
observed born again timescale, the convective mixing velocity ($v_{\rm
MLT}$) in the helium burning shell should be substantially smaller
than that given by the standard mixing length theory of convection
(MLT).  In fact, Herwig (2001) finds that a reduction in the MLT
mixing diffusion coefficient $D$ ($D=\frac{1}{3} \, \alpha_{\rm MLT}
\, v_{\rm MLT}$) by a factor 100 is required to arrive at an agreement
with observations (see also Lawlor \& MacDonald 2003 and Herwig 2002
for a similar conclusion).  The reason for the discrepancy between the
theoretical born again time scales is not clear.

In this paper we present new detailed evolutionary calculations for
the born-again phase by using an independent stellar code that allows
us to follow in detail the abundance changes that take place
throughout all the VLTP phase.  Specifically, we analyze the VLTP
evolution of 0.5842 and 0.5885 \msun\ remnant stars, the previous
evolution of which has been carefully followed from initially 2.5 and
2.7 $M_\odot$ ZAMS star models to the thermally pulsing AGB stage and
following mass loss episodes. The value of the stellar mass of our
post-AGB remnant allows us to compare our predictions with the
observations of the Sakurai's object characterized by a stellar mass
value between 0.535 and 0.7 $M_{\odot}$ (Herwig 2001).  We examine the
dependence of our evolutionary time scales and the phase of proton
ingestion on some numerical and physical parameters.  We extend the
scope of the paper by exploring the role played by the molecular
weight gradient ($\nabla \mu$) induced by proton burning by
considering the double-diffusive MLT for fluids with composition
gradients (Grossman \& Taam 1996).  To the best of our knowledge, this
is the first time that this effect is incorporated in the calculation
of a VLTP.  
The following section describes the main physical
inputs of the evolutionary models, in particular the treatment of the
chemical abundance changes in the convective regions of the star.
Also a brief description of the evolution of the stars prior to the
VLTP is given. In addition, we analyze how the physical hypotheses
made in the modeling of the star set restrictions on the possible
time-resolution during the VLTP, and then the effects of a bad
time-resolution during the proton burning are described.  In Sect.  3,
we present our results, particularly regarding the born-again
evolutionary time scales and proton burning.In Sect. 4 we explore the
dependence of the born again time scale on numerical and physical
details. In Sect. 5 we provide a comparison with observations. And
finally, in Sect. 6 we close the paper by making some concluding
remarks.

\section{Input physics and numerical details}

\subsection{The stellar evolution code}

The evolutionary  calculations presented  in this work  have been  carried out
with the  stellar evolution code LPCODE.  A detailed description  of the code,
both in its numerical aspects as well  as in its input physics, has been given
in Althaus et al. (2003) and references therein.

Here, we  only briefly  review the treatment  given to the  chemical abundance
changes, a key point in the  computation of the short-lived VLTP phase and the
ensuing born-again evolution.  Indeed,  such phases of evolution are extremely
fast to  such an extent that  the time-scale of the  nuclear reactions driving
the  evolution  of  the  star  becomes comparable  to  the  convective  mixing
timescale. Accordingly,  the instantaneous mixing approach  usually assumed in
stellar evolution breaks  down and, instead, a more  realistic treatment which
consistently couples nuclear evolution with time-dependent mixing processes is
required. Specifically,  the abundance changes  for all chemical  elements are
described by the set of equations
\begin{equation} \label{ec1}
\left( \frac{d \vec{Y}}{dt} \right) = 
\left( \frac{\partial \vec{Y}}{\partial t} \right)_{\rm nuc} +
\frac{\partial}{\partial M_r} \left[ \left(4\pi r^2 \rho \right)^2 D 
\frac{\partial \vec{Y}}{\partial M_r}\right],  
\end{equation} 
with $\vec{Y}$ being the vector containing the mole fraction of all considered
nuclear species.  Mixing is treated  as a diffusion process which is described
by  the second  term  of  Eq.~(\ref{ec1}) for  a  given appropriate  diffusion
coefficient  (see  below).   The  first  term  in  Eq.~(\ref{ec1})  gives  the
abundance changes  due to thermonuclear  reactions. Changes both  from nuclear
burning  and mixing  are solved  simultaneously and  numerical details  can be
found in Althaus et al. (2003). The present version of LPCODE accounts for the
evolution  of 16  isotopes and  the chemical  changes are  computed  after the
convergence of each stellar model. A more accurate integration of the chemical
changes  is achieved  by dividing  each  evolutionary time  step into  several
chemical time steps.

The treatment of convection deserves some attention because not only
does it determine how heat is transported in non-radiative regions, but
also because it determines the efficiency of the different mixing
process that can occur in the star. In the present investigation, we
have considered two different local theories of convection.  On one
hand, the traditional Mixing Length Theory (MLT) assuming that the
boundaries of the convective regions are given by the Schwarzschild
criterion. In this case, mixing is restricted only to convective
unstable regions (and the adyacent overshooting layers, as
mentioned below). On the other hand, we have also considered the
double diffusive mixing length theory of convection for fluids with
composition gradients developed by Grossman et al. (1993) in its local
approximation as given by Grossman \& Taam (1996) (hereafter GT96).
The advantage of this formulation is that it applies consistently to
convective, semiconvective and salt finger instability regimes.  In
particular, it accounts for the presence of a non-null molecular
weight gradient $\nabla\mu$.  This will allow us to explore the
consequences of the chemical inhomogeneities that develop in the
convection zone resulting from the fast ingestion of protons during
the born-again evolution.   
As mentioned, the GNA theory describes mixing episodes not only at
dynamically unstable regions (Schwarzschild-Ledoux criterion), but
also accounts for ``saltfinger'' and ``semiconvective'' regimes. Both
of these regimes are consequence of a non zero thermal diffusion rate.
In particular saltfinger mixing takes place in those layers with
negative chemical gradient ($\nabla\mu<0$, i.e.  heavier elements are
above lighter ones) and it is a process far less efficient than the ordinary
convective instability (dynamical instability).

Overshooting has been adopted by following the formulation of Herwig et
al.  (1997) and applied to all convective boundaries in all our
evolutionary calculations.  The adopted overshooting parameter $f$ is
0.016 unless otherwise stated, because this value accounts for the
observed width of the main sequence and abundances in
H-deficient, post-AGB objects (Herwig et al.  1997, Herwig et
al.  1999, Althaus et al. 2005). It should be noted, however, that
there is no physical basis so far supporting a unique choice of $f$
for so different situations in which convection is present in stars.

Before  presenting  our  evolutionary  calculations,  let  us  discuss  a  few
numerical   issues.    Their  importance   will   be   made   clearer   in
Sect.~\ref{sec:alternatives}, where results are presented.

\subsection{Time Resolution during the ingestion of protons}
\label{sec:numerics}

Evolution during  the VLTP  is characterized by  extremely fast changes  in the
structure of the star. This  is particularly true regarding the development of
the H  burning flash when protons  from the stellar  envelope are mixed
into the hot helium-burning shell. 
As was  noted by  Herwig (2001) the  minimum time  step (MTS) used  during the
proton  burning has  to be  kept consistent  with the  hypotheses made  in the
modeling of the star. In particular this means that the MTS must remain larger
than the hydrodynamical time scale, that is of the order of
\begin{equation}
  \tau_{\rm hydr} \sim \sqrt{\frac{R^3}{G M}}.
\end{equation}
At the onset of the VLTP and during the ingestion
of  protons  ($R=2.36  \times   10^9$\mbox{cm})  $\tau_{\rm  hydr}$  is  about
12.9\mbox{s}.  In addition, we have checked that the hydrostatic approximation
is fulfilled to a high degree, as
\begin{equation}
\frac{1}{4  \pi  r^2}  \frac{\partial  ^2 r}{\partial  t^2}  \lesssim  10^{-4}
\frac{Gm}{4 \pi r^4}.
\end{equation}
 in every mesh point during  the H flash and the first
years after it.  Also, the MTS has to be kept larger than the
convective turn-over time (i.e.  the time for the temperature gradient
to reach the values predicted by the mixing length theory, or any
other time-independent convection theory). This time scale can be
estimated as the inverse of the Br\"unt-V\"ais\"al\"a frequency, which
yields the time it takes a bubble to change its temperature in a
convectively unstable zone (Hansen \& Kawaler 1994), that is
\begin{equation}
  \tau_{\rm   conv}   \sim  \frac{1}{|   N|}   =\sqrt{\frac{kT}{\mu  m_H   g^2
  |\nabla-\nabla_{\rm ad}| } }.
\end{equation}
At the conditions of proton burning during  the VLTP ( $T \sim 10^8 K$, $g\sim
1.5  \times  10^7  cm/s^2$   and  $(\nabla-\nabla_{\rm  ad})  \sim  10^{-3}$),
$\tau_{\rm conv} \sim  4.6$ minutes. This value depends  somewhat on the exact
position of the proton burning zone, but it is always of only a few minutes.

In order to be consitent with the assumed hypotheses we should use MTS greater
than  $10^{-5}$\mbox{yr}  (i.e. greater  than  5  minutes).  This could  be
problematic  if the  results for  the born  again time-scale  would  present a
dependence on  the election  of MTS  during the proton  burning phase,  (to be
described  in  the   following  section)  even  for  smaller   values  of  the
MTS. Fortunately,  as will be  clear later, all  our theoretical
sequences  computed  with  ${\rm  MTS} <5  \times  10^{-5}$\mbox{yr}  yield
born-again time-scales  that are, for  practical purposes, independent  of the
MTS value, while being consistent with the fundamental hypothesis. 

 \subsection{Errors due to bad time-resolution during the peak of proton 
burning} 
\label{sec:timeres}

We have computed various VLTP sequences with different time
resolutions (i.e.  by setting different values of MTS) during the
vigorous proton burning episode.  In order to get an error estimation
in the evaluation of $L_{\rm H}$, the proton-burning luminosity
(Herwig 2001 and also Schlattl et al. 2001), we have calculated its
value at each time step from the H mass burned in that time interval
(let us denote it $L{\rm^X_H}$), and compared it to the value
integrated in the structure equations ($L{\rm ^S_H}$).

Chemical changes are followed with a smaller time step than the one
used for the structure equations (typically 3-5 times smaller).  As a
result, $L{\rm ^X_H}$ would be expected to be closer to the actual
value of proton-burning luminosity than $L{\rm ^S_H}$.  However, it
should be noted that due to the fact that the chemical changes are
calculated using a fixed stellar structure, during the onset of H
burning, when the outer border of the convective zone penetrates into
more H-rich layers, the value of $L{\rm ^X_H}$ will be lower than it
should be if the advance of the convective zone $during$ the
calculation of the chemical changes were taken into account \footnote{
 For example Stancliffe et al. (2004) find much stronger He burning
luminosities during the thermal pulses at the AGB if structural and
composition equations are solved $simultaneously$ }.  Thus, the
actual value of $L_H$ may be, for small MTS, slightly greater than
that estimated from $L{\rm ^X_H}$ (see Schlattl et al. 2001 for a
similar consideration).

\begin{figure}[h]
\begin{center}
  \includegraphics[clip,width=8.5cm,angle=0]{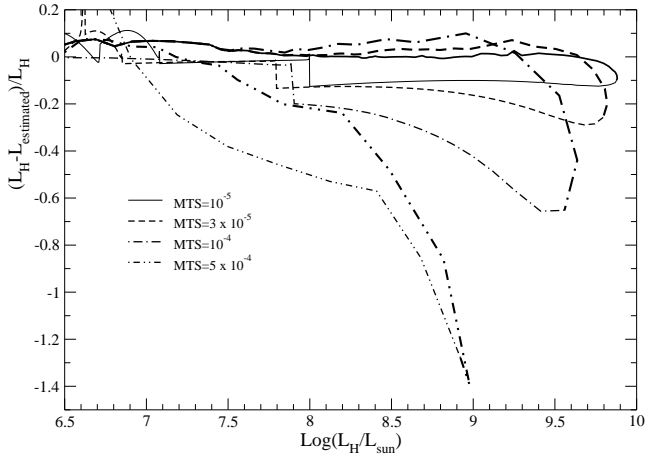}
  \caption{Estimation of the errors involved in $L_H$ (for different
  MTS) during the peak of proton burning for the different models
  calculated with the MLT. Thick lines correspond to models previous
  to the maximum of proton burning, while thin lines show the errors
  after the maximum of $L_H$ }
\label{fig:error_MLT} 
\end{center}
\end{figure}

We  find that  for MTS  smaller than  $5 \times  10^{-5}$\mbox{yr}  ($3 \times
10^{-5}$,$ 10^{-5}$ and $ 3  \times 10^{-6}$\mbox{yr}) the porcentual error in
$L_H$ is always below $10\%$ during  the onset of proton burning. Note that by
using a MTS of $ 1 \times 10^{-5}$\mbox{yr} the error is kept lower than $3\%$
during the onset of  proton burning. On the other hand, when  we use values of
MTS greater than  $10^{-4}$\mbox{yr} the error grows to $50\%$  and even up to
$100\%$  for an  MTS of  $10^{-3}$\mbox{yr}.  In  Fig.~\ref{fig:error_MLT} the
estimated error  of $L_{\rm H}$ during  proton burning is  shown for sequences
with   different  MTS  values. 

As it will be shown in
Sect.~\ref{sec:prot_ing},  the value  of
$L{\rm _H}$  and the  size of  the convective zone  powered by  proton burning
(proton-burning  convection zone,  PBCZ) are  expected to  be  closely related
because an  increase in the energy  flux tends to  promote instability against
convection even further  out in the H-rich envelope,  while an increase
in the PBCZ extension into  the H-rich envelope will increase the value
of $L_H$ as more protons will  become available for being burned at the bottom
of the  convection zone  (Iben et  al. 1983, Schlattl  et al.2001).   Thus, we
expect this  process to be  particularly sensitive to  errors in the  value of
$L{\rm ^S_H}$ at each time step.

\begin{figure*}[t]
\begin{center}
  \includegraphics[clip,height=12cm,angle=0]{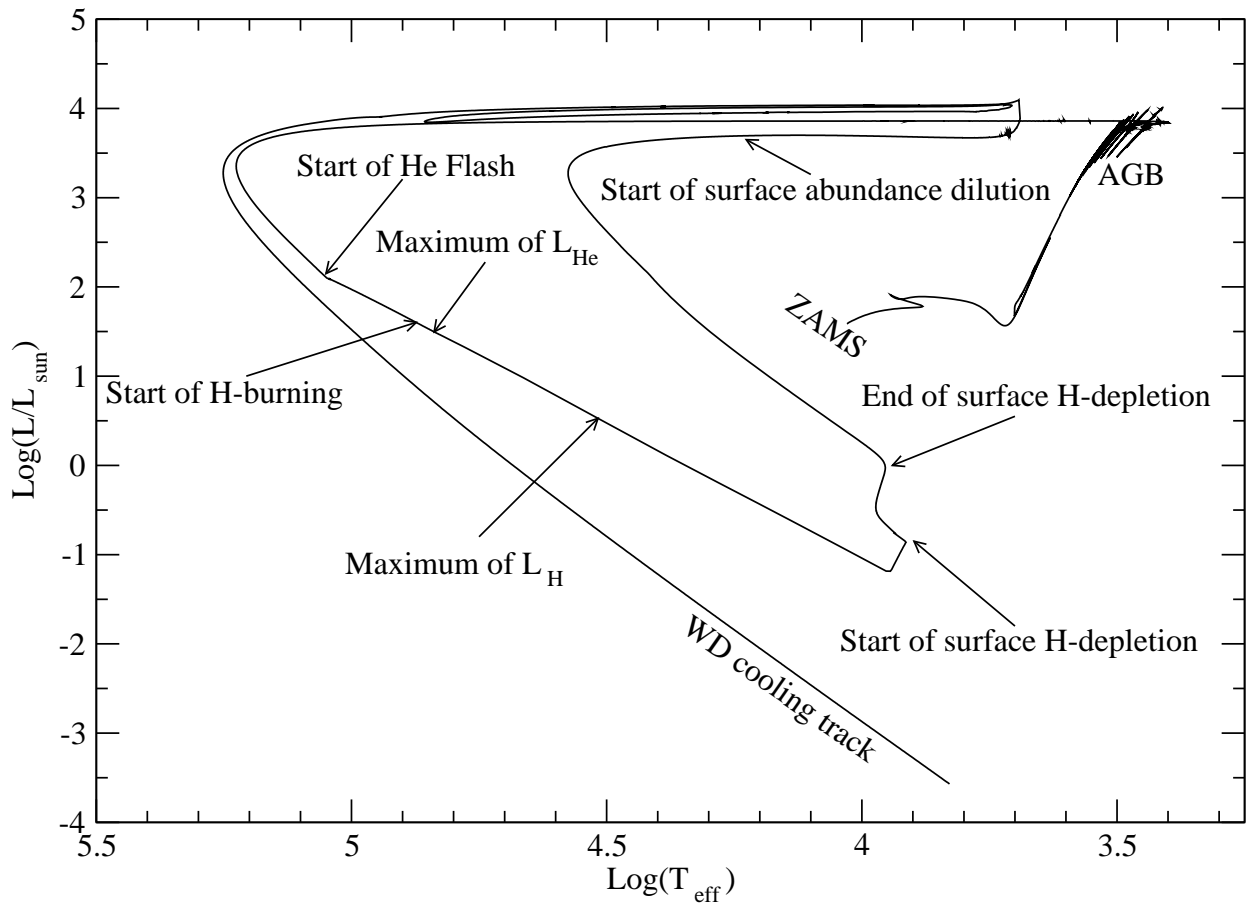}
  \caption{Evolutionary track for the complete evolution of an
    initially 2.5 \msun\ star model from the ZAMS through the AGB, a
    very late thermal pulse (VLTP) to the start of the cooling branch
    as a 0.5842 \msun\ hot white dwarf. Note the occurrence of the
    VLTP at high effective temperature when the post-AGB remnant has
    almost reached the cooling branch.  Relevant episodes during and
    after the VLTP are pointed out. }
\label{fig:HR_full} 
\end{center}
\end{figure*}

\section{Evolutionary results}
\label{sec:results}

\subsection{Prior evolution}

For consistency purposes, the initial VLTP stellar configurations
employed in this study are the result of the complete evolution of
progenitor stars, starting originally from the ZAMS and going through
the thermally pulsing AGB phase.  Unfortunately, the considerable
amount of computing time demanded by our self-consistent solution of
nuclear evolution and time-dependent mixing, has restricted ourserlves
to examining only two cases for the complete evolution for the
progenitor star.  Specifically, the evolution of an initially 2.5
\msun\ stellar model was computed from the zero-age main sequence all
the way from the stages of H and helium burning in the core up to the
tip of the AGB where helium thermal pulses occur. In addition we
compute the full evolution of an initially 2.7 \msun\ stellar model
(see Althaus et al. 2005 for details) which experiences a VLTP
somewhat earlier than the 2.5 \msun\ sequence. The stellar masses of
the post AGB remnants are 0.5842 and 0.5885 \msun\ for the 2.5 and 2.7
\msun\ respectively. A total of 10 and 12 thermal pulses on the AGB
have been calculated.  At the moment of the born again the total
H-mass of the models is $8.5\times10^{-5}$ and  $6.3\times10^{-5}$ \msun.

For both sequences overshoot episodes taking place during central
burning stages as well as during the thermally pulsing AGB phase have
been considered.  A solar-like initial composition $(Y,Z)=
(0.275,0.02)$ has been adopted. 
The complete Hertzsprung-Russell diagram for the 2.5 $M_{\sun}$ sequence
is illustrated in Fig.2.  The evolutionary stages for the progenitor
star (on the cool side of effective temperatures) starting from the
ZAMS as well as the VLTP-induced born again episode and the post born
again evolution towards the white dwarf regime are clearly visible.
Relevant episodes during the VLTP and the ingestion of protons are
indicated as well.  As a result of mass losses the stellar mass is
reduced to 0.5842 \msun. After the born again episode and before the
domain of the central stars of planetary nebulae at high effective
temperatures is reached, the now H-deficient remnant evolves
through a double loop path. Mass loss episodes during and after the
VLTP were not considered.

\subsection{Evolution through the VLTP}

In the rest of this section, we restrict the presentation of
evolutionary results to the 2.5 \msun\ (0.5842 \msun\
remnant) sequence, computed with MTS= $1 \times 10^{-5}$\mbox{yr} and
with the MLT.

\subsubsection{Evolution during the ingestion of protons}
\label{sec:prot_ing}


\begin{figure}[t!]
\begin{center}
  \includegraphics[clip,height=13cm,angle=0]{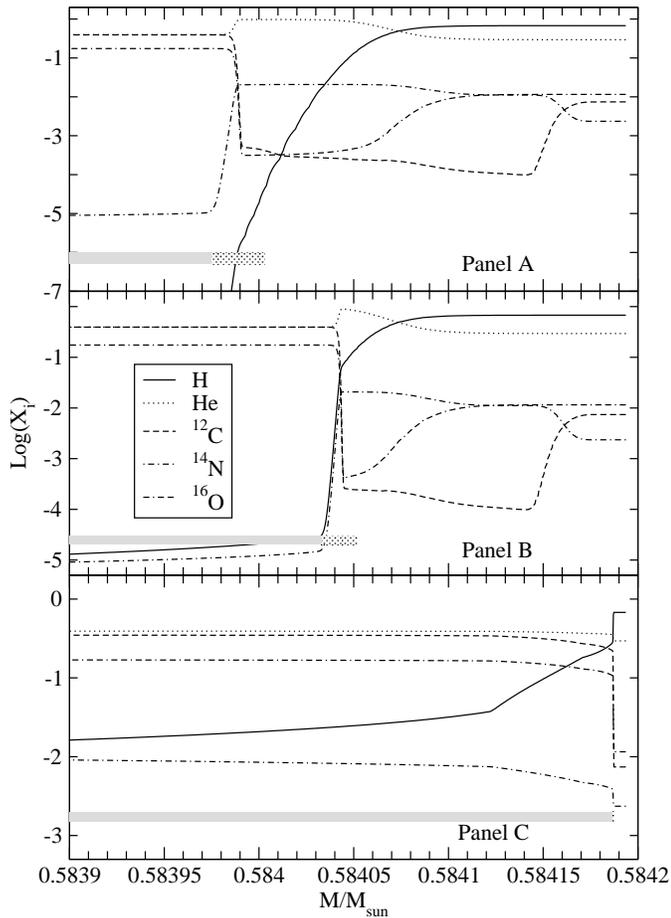} 
\caption{ Evolution of the outer convective border of the PBCZ and the
  inner chemical abundances as a function of mass coordinate during
  the ingestion of protons through the VLTP. Grey bars denote
  convective regions, and dotted bars mark overshoot regions. Panel A
  shows the interior abundances at the moment the He-driven convective
  zone reaches the region where H has been partially burned through
  CNO reactions before the VLTP.  Panel B displays the interior
  compositions at the moment of the splitting of the convective
  region, when convection has reached layers with
  $X\sim0.08$. Finally, panel C corresponds to the moment of maximum
  H-burning energy release.}
\label{fig:ingestion} 
\end{center}
\end{figure}

\begin{figure}[h!]
\begin{center}
\includegraphics[clip,height=7cm,width=8.5cm,angle=0]{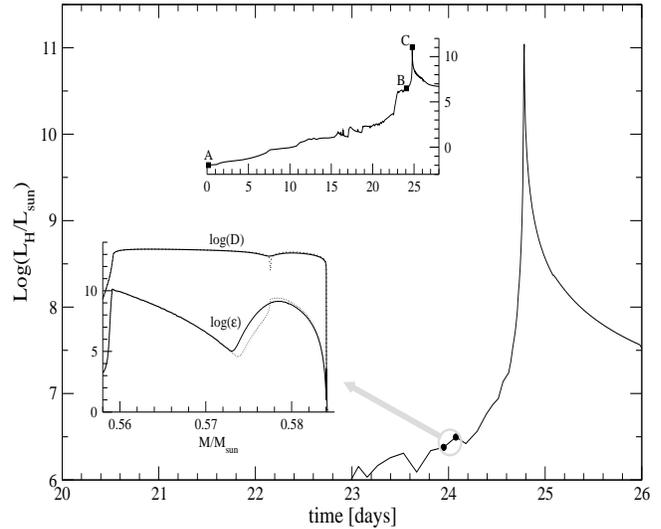} 
\caption{Evolution of $L_{\rm H}$ during the born-again phase (black
dots in the upper inset mark the position of panels A, B, and C in
Fig.~\ref{fig:ingestion}).  The lower inset shows the diffusion
coefficient ($\log{D}$ in \mbox{cm$^2$s$^{-1}$}) and the energy
release ($\log{\epsilon}$ in \mbox{erg g$^{-1}$ s$^{-1}$}), at the
moment just before (continuous lines) and after (broken lines) the
splitting of the original convective region (which causes the drop in
$D$ at M/\msun $\sim$ 0.577) into a He-burning-driven convective zone
and an overlying H-burning-driven convective zone.}
\label{fig:times} 
\end{center}
\end{figure}

During the VLTP, the convection zone powered by the helium shell
burning grows in mass until its outer edge reaches the H-rich
envelope, at variance with the situation encountered in AGB models
where the presence of a high entropy barrier due to H burning prevents
this from occurring. This is shown in panel A in
Fig.~\ref{fig:ingestion}.  It must be noted that the composition at
the base of the H envelope is not homogenous because it is the result
of previous H burning (keep in mind that stars experiencing a VLTP
must have left the AGB during the phase of H-shell burning).  As this
shell (whose width is of 0.00015 ${\rm M_\odot}$) has a H abundance
with a marked depth dependence (see Fig.~\ref{fig:ingestion}), proton
burning rises as layers with a larger amount of H are reached
by convection.  When the outer border of the convection zone reaches
layers where  $X_H \sim 0.035$, $\log{\left(
L{\rm_H/L_\odot}\right)}$ becomes as high as 3. About 2.8 days later,
the convective shell has reached regions with $X_H \sim 0.08$ and
$\log{\left( L{\rm_H/L_\odot}\right)}$ has risen to $\sim 6.5$ (panel
B).  Temporal evolution of $L_{\rm H}$ can be followed in
Fig.~\ref{fig:times}, where also models corresponding to A, B and
C panels of Fig.~\ref{fig:ingestion} are marked in the upper inset.
At this point, two important things take place. First, as a
consequence of the large amount of energy generated in the H burning
shell, the original convective region splits into two distinct
convective zones connected by an overshooting intershell region (see
lower inset of Fig.~\ref{fig:times}). The upper convection zone is the
proton-burning convection zone (PBCZ), while the lower one is powered
by helium burning.  Second, a sudden outwards excursion of the
PBCZ causes an enhanced transport of material from the unprocessed
H-rich envelope downwards to hotter layers where protons are
vigorously burnt, with the consequence that the PBCZ is pushed even
further out.  This second stage of proton burning is extremely violent
and short-lived. In fact, in only  15.4 hours $\log{\left(
L{\rm_H/L_\odot}\right)}$ increases from 6.5 to more than 11. The
sudden increase in the convective region can be seen by comparing the
location of the outer border of the PBCZ in panels  B and C of
Fig.~\ref{fig:ingestion}, where the last one corresponds to the
maximum in $L_H$, as shown in Fig.~\ref{fig:times}.  According to this
description proton burning during the VLTP can be separated into two
stages (a similar situation has been reported by Schlattl et al. 2001
during the ingestion of protons in the core He flash of low
metallicity stars)
The first one lasts for about 25 days while $\log{\left(
L{\rm_H/L_\odot}\right)} \lesssim 6.5$ and the outer convective border
is pushed forward mostly by He-burning. The second stage occurs when
the energy liberated by H-burning becomes comparable to the He-burning
energy release, the convective zone splits into two and the newly formed
PBCZ is pushed outwards by proton burning (and lasts for only a
few hours).  As will be shown in Table~\ref{tab:error}, this feedback
between the increase in the size of the PBCZ and the value of $L_H$ is
expected to make that moderate initial errors in the calculation of
$L_H$ will enhance and produce very different integrated values of
$L_H$ during the born-again evolution, yielding different maximum
values for $L_H$ as well.  Indeed, it is this stage of proton burning
which is seriously altered when we use large MTS, see discussion in
Sect.~\ref{sec:alternatives}.

Finally, let us say that the  violent proton burning occurs
initially mainly through the chain $^{12}$C$+p\rightarrow ^{13}{\rm N}
+ \gamma \rightarrow ^{13}{\rm C} + e^+ + \nu_e $ due to the {\it
extremely} high $^{12}{\rm C}$ abundance in the He-shell.  This
changes as more $^{12}{\rm C}$ is processed into $^{13}{\rm C}$ and
the reaction $^{13}{\rm C} + p\rightarrow ^{14}{\rm N} + \gamma$
becomes more important. At the end of the ingestion of protons both
reactions are generating energy at comparable rates.

\begin{figure}[t]
\begin{center}
  \includegraphics[clip,height=7cm,width=8.5cm,angle=0]{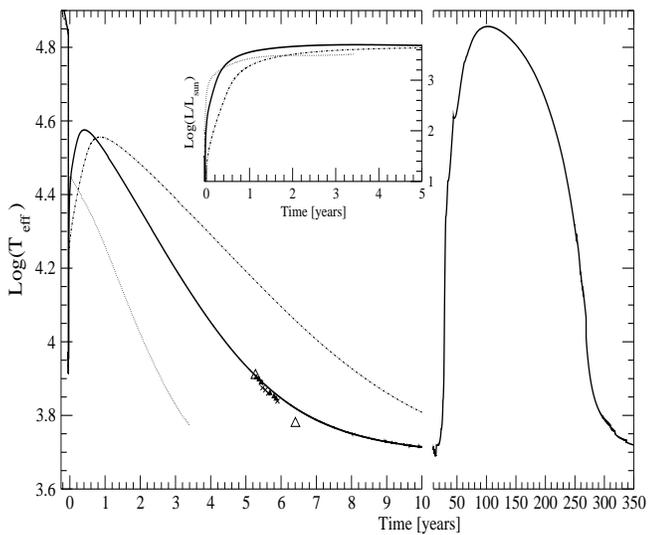} 
\caption{ Evolution of the surface temperature for the sequence
described in section 3 (continuous lines). The zero point of the x-axis
of the observations has been arbitrarily fixed in order to allow for
comparison with the theoretical cooling rate. Triangles
mark the data derived by Duerbeck et al. (1997) (empty triangles
correspond to less reliable data), while crosses correspond to Asplund et
al. (1999). Note the short born again timescale (about 5 years)
displayed by the model. Also the evolution of the effective
temperature for a model which suffers from an ``early'' VLTP (section
4.1, dash dotted line) and for a model with a slight reduction of the
mixing efficiency ($D$/3, section 4.5, dotted line) are shown. Inset
shows the evolution of the lightcurves for the previous models.}
\label{fig:BAtimesk} 
\end{center}
\end{figure}

\subsection{Further Evolution}

\begin{figure}[ht!]
\begin{center}
\includegraphics[clip,height=7cm, width=8.5cm,angle=0]{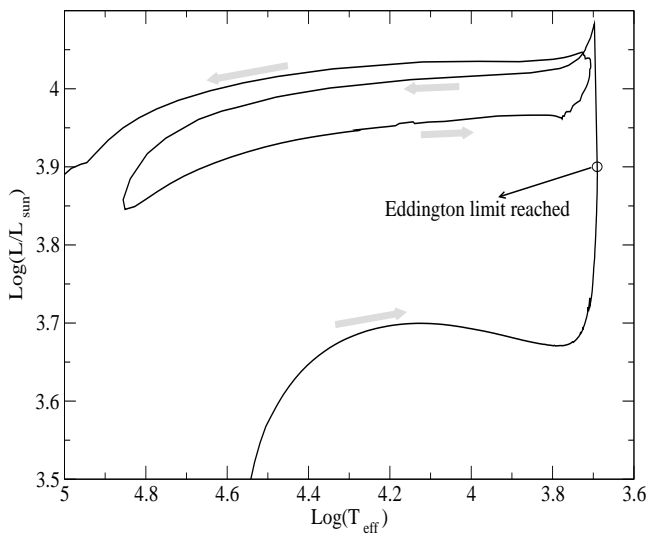}
  \caption{HR diagram for the evolution after the VLTP. Here we see the
  typical double loop pattern due to sucesive H and He driven
  expansions. Also the location where
  our model reaches the Eddington limit is shown.}
\label{fig:doubleloop} 
\end{center}
\end{figure}

After the occurrence of the He-flash and the subsequent violent proton
burning the model expands and returns to the giant region of the HR
diagram in only about 5 yr, which is of the order of the observed
timescale of the Sakurai's object and V605
Aql. Fig.~\ref{fig:BAtimesk} shows the evolution of $T_{\rm eff}$. It
is important to note here that the short evolutionary time-scales are
obtained without invoking a reduction in the mixing efficiency, as
claimed necessary in some previous works (Herwig 2001 and Lawlor \&
MacDonald 2003). The star remains in the red giant domain without
changing its effective temperature significantly ( log$T_{\rm
eff}\lesssim 3.7$) for the next 26 yrs. Although this is at variance
with the reported quick reheating of Sakurai's object (Hajduk 2005),
for log$T_{\rm eff}$ below 3.8 we do not expect our models to be a
good description of what is really happening. This could be because
modeling hypothesis, in particular the hydrostatical approximation,
are not fulfilled in the outer shells of the model \footnote{ Also
it is expected that the H-driven expansion may depend on the mass of
the model. Explicitly, the authors have found that in a 0.664 \msun\
model the H-driven expansion stops at log$T_{\rm eff}=4.8$.}. During
this stage the surface luminosity rises and the layers at the bottom
of the convective envelope attain the conditions where the value of
the radiation pressure becomes high enough to balance the effect of
gravity (Eddington limit, see figure \ref{fig:doubleloop}). When this
happens gas pressure tends to zero, and hydrostatic equilibrium
hypothesis is broken there. This causes our calculations to become
unstable (see also Lawlor \& MacDonald 2003 in one of their
sequences). The Eddington limit provides a natural mechanism to
separate the envelope from the rest of the star (see Faulkner \& Wood
1985).  To proceed with the calculations, we artificially force the
outermost envelope to be in thermal equilibrium (i.e. $dS/dt=0$), this
was done by extending the envelope integration downwards to shells
where the Eddington limit was reached (this happened 15 yr after the H
flash). We note that by doing this we do not alter the surface
abundances of the star (nor its total mass), as the chemical profile
has been previously homogeinized by convection throughout the
envelope. In the next five years the remnant star increases its
surface temperature from $\sim 3.8$ to $\sim 4.35$, and in only 20
$yr$ its temperature increases up to $\sim 4.6$
(Fig.~\ref{fig:BAtimesk}). However these times appear to be dependent
on the detailed treatment of the outer shell of the envelope. About
100 $yr$ after the H flash the star reaches temperatures above
log$T_{\rm eff}=4.8$ and starts the He driven expansion which ends at
log$T_{\rm eff}\sim 3.7$ and lasts for about 250 $yr$ (this is 350
$yr$ after the H flash). Finally after reaching the giant region for
the second time the remnant contracts again and moves to the PG1159
domain. These comings and goings in $T_{\rm eff}$ caused by H and
He-burning driven expansions produce the typical double loop in the HR
diagram (see Fig.  \ref{fig:doubleloop} and also Lawlor \& MacDonald
2003 and Herwig 2003).

\section{Born-again time-scales: dependence on numerical and physical details }
\label{sec:alternatives}

In this section we discuss a variety of issues, both on numerics and physics,
that  may   affect  the   computed  evolutionary  time-scales   of  born-again
episodes. The  reader not  interested in these  details can refer  directly to
Sect.~\ref{sec:discussion}  where  discussion of  the  results and  comparison
with observations are presented.

 As mentioned early in the introduction, theoretical models of the
born again phenomena do not agree on the predicted born again times
for a $\sim 0.6 \rm M _\odot$ star. These times range from about
$\sim$17 yr in Iben \& MacDonald (1995) to 350 yr as in Herwig (1999)
and Lawlor \& MacDonald (2003). In addition the extremely fast
evolution of Sakurai's object has posed a challenge to stellar
evolutionary calculations, which have been unable to reproduce the
observed short timescale unless a high reduction in the mixing
efficiency (of about a factor of 100-1000, Herwig 2001 and
Lawlor \& MacDonald 2003) is assumed. On the contrary, our evolutionary
calculations are able to reproduce born again times of about 5 yr (closer to the observed ones) $without$ changing the mixing
efficiency of the models (see Fig. ~\ref{fig:BAtimesk}). In fact we
are able to reproduce born again times as short as 3 yr by invoking a
reduction in the mixing efficiency by only a factor of 3. We note that
our stellar masses are somewhat smaller than the 0.604 \msun\ used in
Herwig (2001). Although this mass difference is expected to reduce the
born again times (Herwig 2001), it cannot account for the whole
difference in the timescales. In this connection note that the low
mass model (0.535 \msun) presented by Herwig (2001) displays ages still
larger than those derived from our models.

  In view of the different results obtained by different
research groups, we think it is worth investigating how the born-again
time-scale depends on different details entering the calculations,
i.e.  numerics and input physics. In this section we try to address
some of these issues.

\subsection{Results for an early-VLTP sequence}

\begin{figure}[t]
\begin{center}
  \includegraphics[clip,height=8cm,width=9.cm,angle=0]{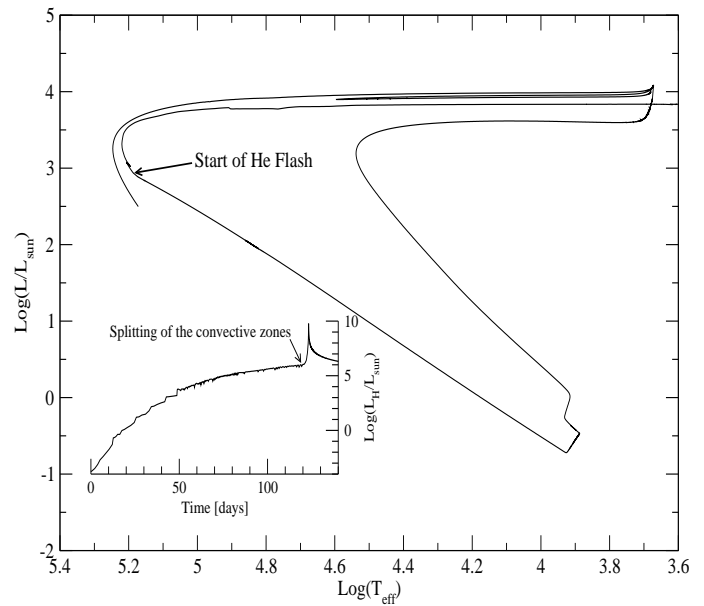} 
\caption{ Same as Fig. 2 but for a model which suffers from an
``early'' VLTP (this is at luminosities about one order of magnitude
larger than in the model shown in Fig. 2).The inset shows the temporal
evolution of $L_H$ for this model. Note that the evolution is less
violent than in the previous case (compare upper inset of Fig. 4) and
that the two stages of proton burning are more easily differentiated.}
\label{fig:viejo-HR-LH} 
\end{center}
\end{figure}

Here we explore the expectations for the 0.5885
\msun\ model that experiences a VLTP earlier (i.e. at higher
luminosities, log$(L/L_\odot)\sim 3$) than the 2.5 \msun\ sequence
\footnote{At comparing the predictions of the two sequences we have to
keep in mind that the stellar masses of the remnants are not strictly
the same.}.  At the moment of the VLTP the object is less compact and
the flash takes place in a less degenerate environment, as compared
with the 2.5 \msun\ sequence. Also note that the total H-mass in the
model (which is almost completely burned during the VLTP) is lower in
this case. Therefore, the proton ingestion described in section 3.2.1
is less violent. As can be seen in the inset of
Fig. \ref{fig:viejo-HR-LH} the two stages of proton burning are more
distinguishable than in the previous case. Longer times are needed for
the H-flash to develope. In this case the first stage of proton
burning (the one before the splitting of the convective regions, when
the outer convective border is driven mainly by He-burning) lasts for
about 0.3 yr, while the second more violent stage takes about 2.5
days. Also note that the magnitude of $L_H$ at the maximum is lower
than in the previous model, reaching a value of log$(L_H/L_\odot)\sim
9.8$. Consequently we find that born again times are longer in this
case, taking about 8.7 yr (see Fig. 5 and 8) when a standard MTS of
$10^{-5}$ yr is adopted during the H-flash (Table 1).

\subsection{Time resolution} 
In this, and in the following sections, the results correspond to the
2.7 \msun\ sequence.

\begin{table}[ht!] 
\begin{tabular}{cccc}\hline
  MTS & $\int{L{\rm_H dt}} $ [erg] & $\Delta t$ [days] & Born-again \\ 
  & & $L_H>(e^{-1} \times {L_H}_{\rm max} )$& time-scale [yr] \\ 
  \hline $1\times
  10^{-5}$ & $4.5\times 10^{47}$ & 0.113 & 8.7 \\ 
  $3\times  10^{-5}$  &
  $4.31\times 10^{47}$ & 0.133 & 10.3 \\ 
  $1\times  10^{-4}$ &  $3.75\times
  10^{47}$ & 0.209 & 21 \\ 
  $5\times 10^{-4}$ & $2.59\times 10^{47}$ & 0.777 &
  280 \\ \hline
\end{tabular} 
\caption{Characteristic values for the VLTP, for various MTS values.
Note the dispersion in born again times according to the adopted MTS
value. Third column shows a characteristic time for the duration of
the peak of proton burning. The born-again times correspond to the
time interval from the moment of maximum $T_{\rm eff}$ after the H
burning (and at high luminosity) until $\log{T_{\rm eff}}=3.8$ is
reached for the first time.}
\label{tab:error} 
\end{table}

In Sect.~\ref{sec:numerics} we discussed the constraints imposed on
the MTS choice by basic assumptions implicit in stellar evolutionary
calculations, namely those of hydrostatic equilibrium and the use of
time-independent theory of convection. Then in Sect
\ref{sec:timeres} it was noted that the chemical integration scheme
used in this work was sensitive to the adopted time resolution. Now we
analyze how a poor time resolution affects the energy liberation and,
consequently, the born again times.

\begin{figure}[ht]
\begin{center}
  \includegraphics[clip,height=8cm,width=9cm,angle=0]{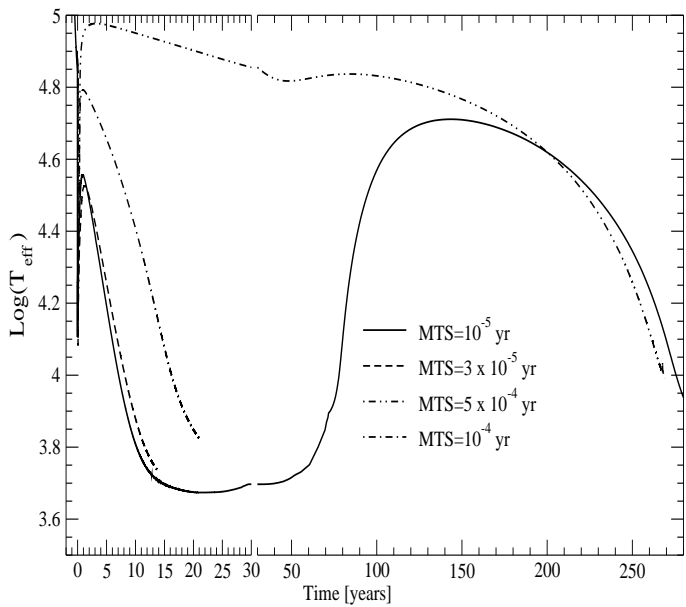} 
\caption{ Evolution of the surface temperature for models
calculated with different MTS. Note the high dependence of the born
again times with the chosen time resolution during the H-flash when
the MTS is above $5\times10^{-5}$ yr. In particular note that for a
very poor time resolution (MTS$\sim 5\times 10^{-4}$ yr) the H-driven
expansion is almost totally suppressed. }
\label{fig:BAtimestimeres} 
\end{center}
\end{figure}

\begin{figure}[t]
\begin{center}
  \includegraphics[clip, , height=6cm,
  width=8.5cm,angle=0]{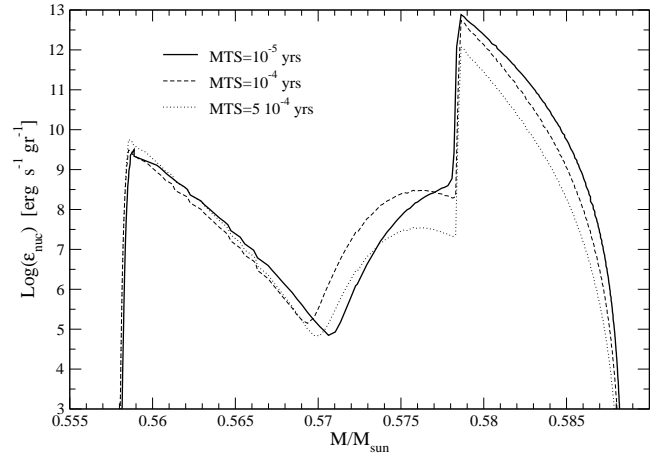}
\caption{ Nuclear energy generation due to helium and proton burning
as a function of the mass coordinate for various MTS values. The most
external peak shown corresponds to the energy released by proton
burning. Note the change in the energy release resulting from
different time resolution. This is because a good time
resolution leads to a more violent proton ingestion, and thus to a
higher H-burning rate.  Indeed, this caused both a change in the total
energy generated and a change in the interval of time in which it is
liberated (see table \ref{tab:error}). Note also that the location
of the main H burning energy release is the same irrespective of the
election for the MTS }
\label{fig:PBZ_MLT} 
\end{center}
\end{figure}

 In Table \ref{tab:error} we show some characteristic values of the
   violent proton burning for different calculations of the process
   according to the MTS employed.  Note from Fig.~\ref{fig:error_MLT}
   that a poor time resolution yields $L{\rm ^S_H}$ values that are
   underestimated with respect to $L{\rm ^X_H}$ by more than an order
   of magnitude (for the definition of $L{\rm ^S_H}$ and $L{\rm ^X_H}$
   see Section \ref{sec:timeres}). Also note that when evolution is
   calculated using a MTS of $5 \times 10^{-4}$ yr, the energy is
   liberated over a period $\sim7$ times larger, and the total energy
   liberated by proton burning is $40\%$ smaller than in the standard
   case with MTS $=10^{-5}$\mbox{yr}. As a result the born again
   timescale strongly depends on the election of the MTS, as shown in
   Fig.~\ref{fig:BAtimestimeres} (see also Fig. \ref{fig:PBZ_MLT}
   where the rate of energy generation per unit gram, at the peak of
   proton burning, changes one order of magnitude) . A large value of
   the MTS leads to born-again time-scales of the order of
   300~\mbox{yr}, which are typical values corresponding to born-again
   episodes driven by helium-burning (Iben et al., 1983). When time
   resolution is improved and the MTS is chosen to be lower that
   $5\times 10^{-5}$~\mbox{yr} much shorter evolutionary times for the
   born-again event are obtained. As can be seen in
   Fig.~\ref{fig:BAtimestimeres}, it takes only 10~\mbox{yr} for the
   remnant to evolve from the white dwarf configuration to giant
   dimensions.  This dependence can be understood in terms of the
   large errors (between 30\% and 100\%) that lead to an
   underestimation of $L{\rm ^S_H}$ when large MTS are used; while
   taking MTS shorter than $5\times 10^{-5}$~\mbox{yr} keeps these
   errors below the 10\% level (Fig.~\ref{fig:error_MLT} and
   Table~\ref{tab:error}).  It is interesting to note that these
   changes in the born-again time-scales are not related with the
   position of the maximum proton-burning zone in the models, which is
   the same irrespective of the MTS used, as illustrated in
   Fig.~\ref{fig:PBZ_MLT}.  A final important point to be remarked is
   the convergence we find for the evolutionary times, i.e. using MTS
   values smaller than $5\times 10^{-5}$~\mbox{yr} do not lead to
   shorter timescales. This seems to indicate that errors in the
   evaluation of $L{\rm ^S_H}$ are under control and that results for
   the time-scales are robust, at least regarding the time-step
   choice\footnote{We note that for our best model the energy
   liberated by proton burning is in the range of expected values from
   the following rough estimation: If the whole H-mass ($6.3 \times
   10^{-5}$ \msun ) were burned only through the chain
   $^{12}$C$+p\rightarrow ^{13}{\rm N} + \gamma \rightarrow ^{13}{\rm
   C} + e^+ + \nu_e $ (in which each burned proton liberates 3.4573
   Mev, as it happens initially) then $\int{L{\rm_H dt}} $=$4.15\times
   10^{47}$ erg. While if the complete H-mass were burned through the two
   important reactions (the previous one and also $^{13}{\rm C} +
   p\rightarrow ^{14}{\rm N} + \gamma$) working at the same rate (in
   this case 5.504 Mev are liberated per burned proton) then
   $\int{L{\rm_H dt}} $=$6.6\times 10^{47}$ erg.}.

\subsection{Convection theory: Effect of the $\mu$-gradient ($\nabla\mu$)}
In order to investigate the effect of the chemical gradients
($\nabla\mu$) on the mixing efficiency we carried out additional
calculations with the GNA theory. As was mentioned, this theory
accounts for the presence of non-null molecular weight gradients (see
Sect. 2).  Firstly, we compare the GNA with the standard MLT theory. To
avoid differences in the born again times coming from a different
prescription for the relation between the diffusion coefficient and
the convective mixing velocity we adopted for both theories the same
relation $D= \frac{1}{3} \times l \times v$
\footnote{In the previous work by Althaus et al. (2005) the relation
used was $D= l \times v$ and that's (mainly) why they find greater
born again times than in the present work}. The convective velocity
for the MLT ($v_{\rm MLT}$) is taken from Langer et
al. (1985)\footnote{We have also performed some simulations using the
expresion for $D$ derived from Cox \& Giuli (1968) ($D=\alpha^{4/3}
H_P \left[\frac{c}{\kappa \rho} (1-\beta) \nabla_{\rm ad} (\nabla_{\rm
rad}-\nabla)\right]^{1/3}$).  We find that no significant difference
arises from this change.  }  ), while $v_{\rm GNA}$ is taken from
Grossman \& Taam (1996). The mixing length ($l$) is taken to be 1.7 and
1.5 times the pressure scale height for the MLT and GNA respectively. In
order to dissentangle the effect of $\nabla\mu$ in the value of
$D_{\rm GNA}$, we calculated two different sequences: one considering
the standard GNA and the other by setting to zero the value of
$\nabla\mu$. We find, as we expected, that GNA yields similar born
again times as those given by the MLT when $\nabla\mu$ is set to zero.
\begin{figure}[htp]
\begin{center}
  \includegraphics[clip,height=11cm,width=8.5cm,angle=0]{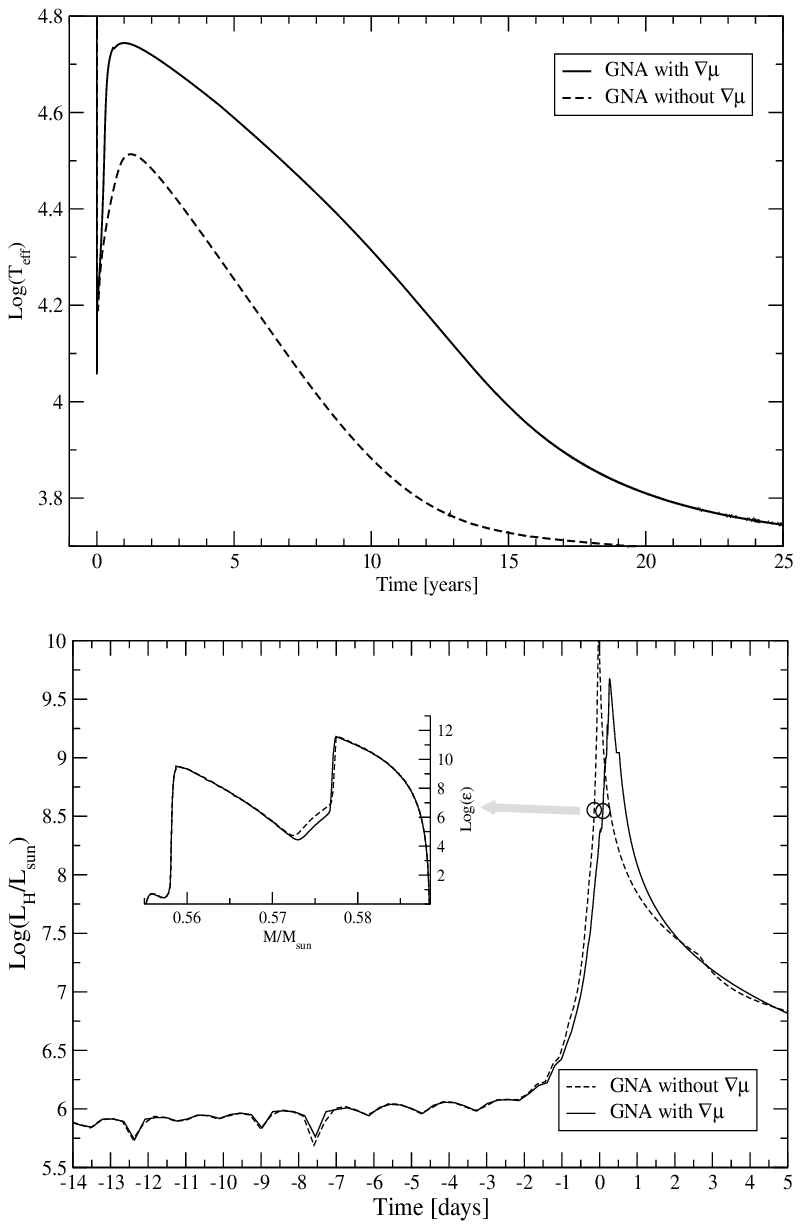} 
\caption{Evolutionary  time-scales during the  born-again event  for sequences
  computed with  the GNA convection theory  with and without the  effects of 
  $\nabla\mu$ included in  the calculation  of the
  mixing coefficient  $D$ (top  panel).  Evolution of  $L_{\rm H}$  during the
  development of the H-flash (bottom panel).  
  The influence of  $\nabla\mu$ becomes noticeable only when  the second stage
  of proton  burning sets in. The  effects of $\nabla\mu$  become important as
  material  with lower  $\mu$  is  rapidly injected  into  the burning  shell,
  slightly delaying the violent increase of $L_{\rm H}$. The
  inset shows the small effect of  the molecular weight gradient in the energy
  release distribution.} 
\label{fig:mu_ages} 
\end{center}
\end{figure}
Next, we analyze the effect of chemical inhomogeinities in the
convective shell during proton burning. To this end we compare the two
sequences calculated with the GNA theory (one including the effect of
$\nabla\mu$, and the other by setting $\nabla\mu=0$ in the calculation
of $D$). A comparison of these sequences is shown in
Fig.~\ref{fig:mu_ages}. During the first stage of the onset of proton
burning, the amount of protons ingested in the convection zone is
relatively small and $\nabla\mu$ is not significant to alter the value
of $D$. Thus, the presence of $\nabla \mu$ does not change the
evolution of this stage of proton burning.  However, during the second
stage of proton burning, larger quantities of H are rapidly
transported downwards, causing $\nabla\mu$ to be
larger\footnote{Because of the very short time during which this
process takes place, the H distribution is not homogeneous.}.  A
positive $\nabla\mu$ tends to favour convective stability, resulting
in a lower value of $D$ than if $\nabla \mu$ is assumed to be zero.
In addition, a smaller value of $D$ implies lower rates of proton
burning, leading to a lower value of ($\nabla_{\rm rad}-\nabla_{\rm
ad}$) which in turn increases the difference between $D$ and
$D(\nabla\mu=0)$. As a result of this slight reduction in $D$ at the
PBCZ, proton burning becomes somewhat less violent and lower values of
$L_H$ are attained (bottom panel of Fig.~\ref{fig:mu_ages}), with the
consequence that the born again times are larger (upper panel of
Fig.~\ref{fig:mu_ages}).  Note that this is not related to the
location of the peak of proton burning, which is not affected by the
inclusion of $\nabla\mu$ (because $\nabla\mu$ is not important until the start of the second stage of proton burning).


\subsection{Effect of the overshooting parameter $f$}

To explore the role of the overshoot parameter $f$, we have
calculated an additional MLT-sequence with $f=0.03$ . It may
be argued that changing $f$ only at the VLTP stage is not consistent
with the prior evolution.  We do not expect, however, this
inconsistency to alter the main aspects of the following
discussion.\footnote{ Changing the efficiency of overshooting during the
thermally pulsing phase is expected to modify the amount of $^{16}$O
in the intershell region below the helium buffer (Herwig 2000).}  In
fact, there is no reason to believe that the overshooting efficiency during
the extreme conditions characterizing the VLTP should remain the same
as that prevailing during prior evolutionary stages.  Interestingly
enough, we find that born-again evolution is sensitive to the adopted
$f$. 
\begin{figure*}[htp]
\begin{center}
\includegraphics[clip,height=10cm,angle=0]{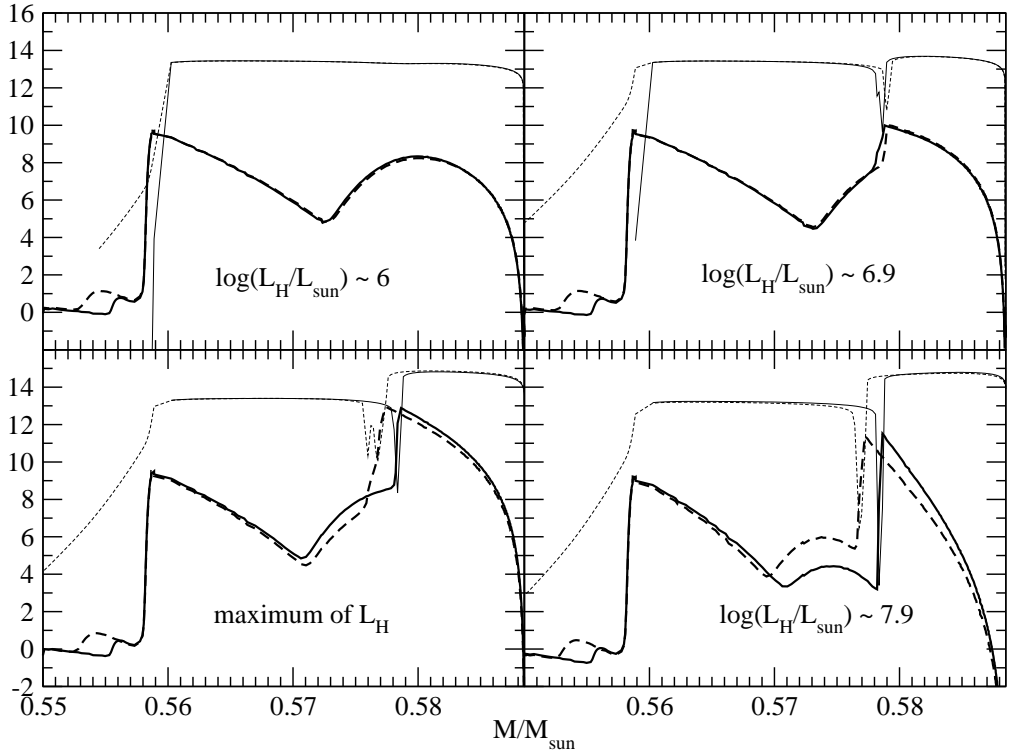} 
 \caption{Location of the peak of the energy liberated by proton
  burning at selected stages during the ingestion of protons for
  different values of the overshooting parameter $f$. Solid lines show
  the diffusion coefficient and the energy liberated (per unit time
  and mass) for the model with $f=0.016$ while dashed lines show the
  situation for $f=0.03$ (thin lines correspond to log$D$ and thick
  lines log$\epsilon$). Upper left: just before the splitting of
  helium flash convection zone as a consecuence of the energy
  generated by proton burning. Upper right: when the convective region
  has just splitted into two. Bottom left: the situation at the maximum
  of $L_H$. Bottom right: after the maximum of proton burning.}
\label{fig:f=0.03}
\end{center}
\end{figure*}
 The sequence with increased overshooting
efficiency takes about 20~\mbox{yr} to reach $\log{T_{\rm eff}} \sim
3.8$, as compared to the 10~\mbox{yr} employed by the $f=0.016$
sequence. The reason for the slower evolution in the $f=0.03$ case can
be understood by examining the evolution of the locus of maximum
proton burning during the ingestion of protons as shown in
Fig.~\ref{fig:f=0.03}.  Note that, while in both sequences the peak of
proton burning (and the moment of the splitting of the convective
zones) starts to develop at about the same position (in mass) at
$m\sim 0.58$~\msun, the situation changes soon. In fact, for the
sequence with $f= 0.016$ the peak of proton burning moves downwards
only to $m\sim 0.579$~\msun\ as the energy released by proton burning
increases, but for the case with $f=0.03$ it progressively sinks into
deeper regions of the star, reaching a mass depth of 0.577 \msun\ by
the time $L_H$ has reached its maximum value.  A larger overshooting efficiency
allows protons to reach deeper layers, below the inner boundary of the
PBCZ. This has the effect of shifting the maximum nuclear
energy release to inner regions that become unstable to convection
because of the high luminosity.  The net effect, as mentioned above,
is a final displacement in the peak of proton burning to deeper regions
of the star by more than 0.003 \msun\ in mass. Correspondingly,
evolutionary time-scales are longer.

 An issue raised by the referee is concerning the possibility that
the overshooting mixing at the H-He interface could be attenuated by
the stabilizing effect of the chemical gradient. To analyze this
possibility we have calculated two more GNA-sequences under the
$extreme$ situation in which no overshooting is present at the outer
border of the He-driven convection zone (OBHeCZ).  Under this
assumption our two sequences display a very different behavior. In
the 2.7 \msun\ sequence, which suffers from an early-VLTP and in which
some entropy barrier is still present, the presence of $\nabla\mu$
prevents the He-driven convection zone from reaching the H-rich
envelope. Thus in this sequence H-burning was inhibited. On the other
hand in the 2.5 \msun\ sequence (in which proton burning is almost
completely extinguished at the moment of the He-flash) the
$\nabla\mu$-barrier is not enough to prevent proton burning. However
evolution proceeds differently from the case in which overshooting
mixing was present. First the starting of proton ingestion was delayed
for about 10 days. Then, once the ingestion of protons has started, as
consequence of $\nabla\mu$ some semiconvective regions develop that
split the OBHeCZ into separate convective regions. Then H is premixed
in the more external convective regions before being ingested and burned. As a
result, log$(L_H/L_\odot)$ remains below 7 for about half
year delaying the start of the second (runaway) stage of proton
burning. Only when $L_H$ has became high enough and the H-abundance in
the pre-mixed convective regions has been lowered are these
convective regions reconnected. Then, due to this change in
the way protons are burned the born again times change, being much
larger (about 100 yr) than in the case in which overshooting was
considered.  

\subsection{Miscellanea}

Our evolutionary results do not indicate the need of a strong
reduction in mixing coefficient $D$ to achieve born-again time-scales
comparable to the rapid evolution showed by the Sakurai's object.
However, a reduction in $D$ certainly leads to faster born-again
evolution. As it is show in Fig.  \ref{fig:BAtimesk} a $slight$
reduction in the mixing efficiency by a factor of 3 shortens the born
again time to about 3 years. 

We have perfomed additional VLTP computations to assess the
possibility that our conclusions could be affected by other numerical
resolution issues.  We find no dependence of our born-again
time-scales on the chemical time step used in the integration of the
chemical evolution. Nor do we find any significant dependence on the
adopted mesh resolution. Also as explained in Appendix A, two
different approaches for the linearization of the structure equations
were used in the calculations, and no significant difference in the
born again times was found. Additionally, we explored the effect of
possible uncertainties (of about 10\% to 20\%) in the rate of the
$^{12}{\rm C}+ p \rightarrow ^{13}{\rm N} + \gamma$. Again, we find no
relevant effect on the born again times.

\section{Comparison with observations and chemical evolution of the models}
\label{sec:discussion}

\subsection{Pre-Outburst parameters}

\begin{figure}[ht!]
\begin{center}
\includegraphics[clip,height=7cm, width=9.cm,angle=0]{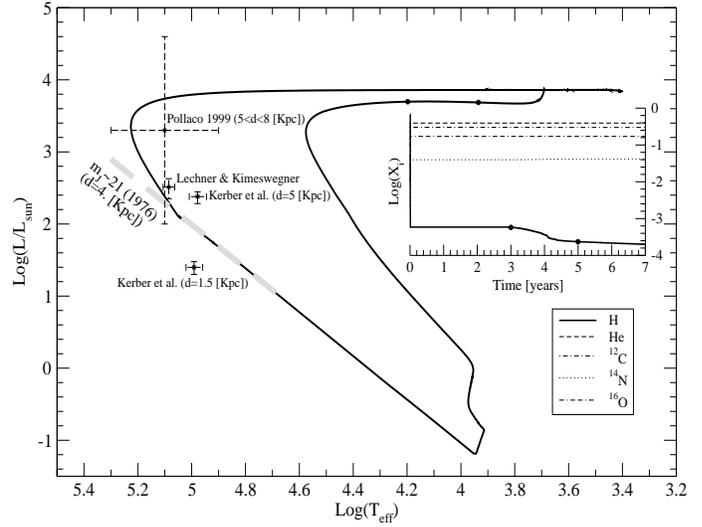}
\caption{ Comparison of our standard model (section 3) with the
preoutburst properties derived from observations. Grey dashed line
corresponds to the possible detection at m$_j$=21 in 1976 (taken from
Herwig 2001), if a distance of 4 Kpc is assumed. Crosses mark the
location of Sakurai's object and V605 Aql previous to the helium flash
as derived from photoionization models (Pollaco 1999 and Kerber et
al. 1999 for Sakurai's object and Lechner \& Kimeswegner 2004 for
V605 Aql. Inset: Evolution of surface abundances. Note the change in
the surface H while the other elements remain unchanged. A similar
behavior in  the surface abundances has been reported in Sakurai's
object during 1996 (Asplund et al. 1999).  Black dots mark the
location in the HR before and after the second change in the
H-abundance. }
\label{fig:surfacequimi} 
\end{center}
\end{figure}

In Fig. \ref{fig:surfacequimi} we compare the HR-locus of the model at
the moment of the He-Flash with that derived from photoionization
models for Sakurai's object and V605 Aql (Pollaco 1999, Kerber et
al. 1999 and Lechner \& Kimeswenger 2004), and with the possible
detection of Sakurai's object as a faint object of $m_J\sim21$ on the
J plate of the ESO/SERC sky survey taken in 1976 (Duerbeck \& Benetti
1996, Herwig 2001). As it is shown in Fig. \ref{fig:surfacequimi} our
model shows a good agreement with the possible detection at
$m_J\sim21$ when a distance of 4 Kpc is assumed. Comparison with
photoinoization models shows a relatively good agreement, in particular
when a distance scale (which determines the zero point of the y-axis
of the observations) between 5 Kpc and 1.5 Kpc is chosen.

\subsection{Evolution of effective temperature and luminosity}


In Fig. \ref{fig:BAtimesk} we show the temporal evolution of the
luminosity and the temperature for both the theoretical models and the
observed parameters of Sakurai's Object (Duerbeck et al. 1997,
Asplund et al.  1999). Although important quantitative disagreement
exists with observations, some of the observed features of
Sakurai's object are predicted by the models. Indeed note that the
cooling rate of the models is similar to the observed one. When
comparing the timescales of V4334 Sgr with those of the models, we
have to keep in mind that the observed born again times are measured
from the moment the object has become bright enough (and not from the
moment of the flash itself). 
The light curve of our models during the first years after the violent
proton burning is roughly similar to the observed one, that is, the
object increases its luminosity by more than 2 orders of magnitude in
less than a year (Fig.\ref{fig:BAtimesk}), in agreement with the
observed one (Duerbeck et al. 1997, Duerbeck et al. 2002).In addition
the log($L/L_{\odot})$ value of 3.5-3.7 predicted by our models after
the outburst is similar to the luminosity derived for V4334 Sgr (by
Duerbeck and Bennetti 1996) when a conservative distance scale of 5.5
kpc is adopted (log$(L/L_{\odot})\sim 3.49$, $T_{\rm
eff}\sim3100$). On the other hand our models are incompatible with the
luminosity of 10000 $L_\odot$ derived when adopting the long distance
scale (8 kpc, Duerbeck et al 1997).

On the other hand our models fail to reproduce the quick reheating
(less than 6 years after reaching log$T_{\rm eff}\sim3.8$) shown by
Sakurai's object as reported by Hajduk et al. (2005). This could be
probably related to the fact that the hydrostatical hypothesis assumed
in our modeling is  explicitly broken in the outer regions of the
star. In this connection let us note that although Hajduk et
al. (2005) claim that this feature can be reproduced by means of
models with a high supression of the mixing efficiency, we think that
this has to be taken with a pinch of salt. The lowest log$(T_{\rm
eff})$ reached by the model presented in that work is 3.95 and thus
can not reproduce the observed evolution of the Sakurai's Object
effective temperature, that has reached values as low as log$(T_{\rm
eff})\sim$3.8.  Also notice that it is below log$(T_{\rm eff})=3.8$
that we find that hydrodynamical effects in the envelope become
important, and ejection of envelope material is likely to occur. In
connection with this, it is worth mentioning that the Sakurai's Object
has been observed to go through a massive dust shell phase when its
effective temperature was below log$(T_{\rm eff})=3.8$ (Duerbeck et
al.  2000).

\begin{figure}[ht!]
\begin{center}
\includegraphics[clip,height=13cm,angle=0]{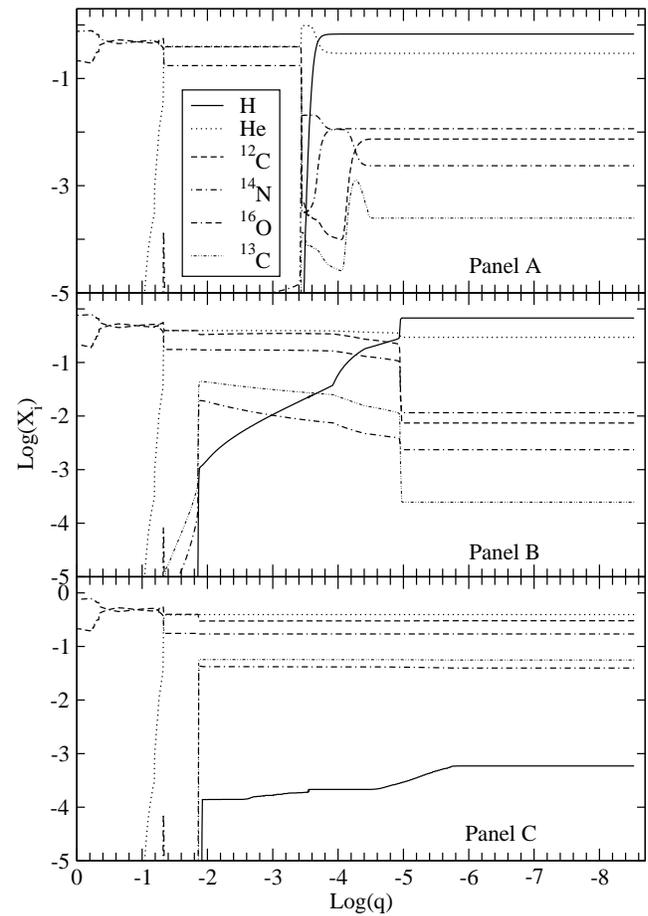}
  \caption{Inner Chemical evolution. Panel A shows the chemical
  profile before the onset of proton ingestion (He burning is located
  at about log$(q)$=-1.37). Note that almost all the previously
  created $^{14}$N has been destroyed. Panel B shows the interior of
  the models at the maximum of proton burning, which has its maximum
  intensity near log$(q)$=-1.88. Note especially the inhomogenous
  distribution of H at the PBCZ (which at that moment ranges from
  -1.88 to -4.9). Also see the creation of $^{13}$C and $^{14}$N due
  to the ingestion of protons.  Finally last panel shows the chemical
  abundances somewhat later when the model start to cool as
  consequence of the expansion of its outer layers. At that moment the
  envelope is almost homogenous as a consequence of convection,
  but the H-abundance left by proton burning displays a marked
  depth dependence.}
\label{fig:chemical} 
\end{center}
\end{figure}

\subsection{Chemical evolution of the models}

In Fig. \ref{fig:chemical} the evolution of the abundance distribution
within the star during the VLTP is shown. Until the ingestion of
protons, the chemical abundance distribution is essentially that of
its AGB predecesor. During the VLTP, the outwards growing helium flash
convection zone penetrates into the helium buffer, and as a result the
$^{14}$N left behind by CNO reactions in the helium buffer is burned
into $^{22}$Ne. This leads to an almost total depletion of $^{14}$N in
the He convective zone.  The course of events up to this point is
similar to that occurring during the thermally pulsing AGB phase, but
things change as soon as protons are ingested into the convective
region. Initially the ingestion of protons causes only an increase in
$^{13}$C. This continues until the convection region splits into two
and unstable proton burning develops. A few hours after the splitting
of the convective region, when enough quantities of $^{13}$C have been
produced, the reaction $^{13}$C(p,$\gamma$)$^{14}$N takes over and the
amount of $^{14}$N in the PBCZ increases. Also after proton burning
has reached its maximum extent, $^{14}$N and $^{13}$C, which had
initially a maximum abundance at the place of major proton burning are
soon homogenized by convection in the whole PBCZ (last panel of
Fig. \ref{fig:chemical}). This process leads to an important $^{14}$N
abundance in contrast with what would be expected in a Late Thermal
Pulse (LTP), where most $^{14}$N abundance is destroyed during the He
flash and never recovered.

\begin{table*}[ht!]
\begin{center}
\begin{tabular}{ccccccc}\hline
& H & He & C & N & O & $M^{total}_H/M_\odot$ \\\hline
MLT (early VLTP) & $3.7 \times 10^{-4}$ & 0.324 & 0.371 & 0.0145 & 0.217 & $3.19 \times 10^{-6}$ \\ 
MLT  ($f=0.03$) & $5\times 10^{-6} $ & 0.307 & 0.378 & 0.0129 & 0.229& $7.24 \times 10^{-9}$ \\ 
GNA ($\nabla\mu=0$) & $3.2\times 10^{-5} $ &  0.316 & 0.378 & 0.0129 & 0.221 &$2.19 \times 10^{-7}$ \\ 
GNA &  $3.3\times 10^{-5} $ &  0.316 & 0.378 &  0.0127 & 0.221  & $6.76 \times10^{-9}$ \\\hline 
Lawlor   $D/10^{-4}\dagger$&0.06  &0.57 &0.26  & -  & 0.07  &  $2 \times10^{-6}$\\ 
Lawlor   $D/10^{-3}\dagger$&0.02  &0.55 &0.28  & -  & 0.08  &  $3 \times10^{-7}$\\ 
Lawlor  $D/10^{-2}\ddagger$&0.04 &0.54 &0.28 & - & 0.08 & - \\ 
Herwig(1999)& - & 0.388 & 0.362 & - & 0.219 & $4 \times 10^{-11}$ \\ \hline
Most PG1159 pulsators & - & 0.397  & 0.357 & 0.0139 & 0.159 & - \\ 
PG1159-035            & - & 0.287  & 0.516 & 0.0100 & 0.115 & - \\ \hline 

\end{tabular} 
\caption{Surface abundances of the models presented in
Sect. 4. Abundances are taken at the moment the models reach the giant
region for the first time after the VLTP. Abundances from Lawlor \&
Macdonald (2003) are taken from models of similar metallicity (z=0.02)
$\dagger$ Abundances correspond to the moment at which the star
reaches the WD cooling track. $\ddagger$ Abundances correspond to the
first return to the giant region, where the envelope was homogeinized
by mixing. In the last column the values of the total amount of H in
the star after the VLTP are presented. Relative abundances of PG1159
stars are taken from Dreizler \& Heber (1998) and normalized to the
total amount of C+O+N+He given by the models.}
\label{tab:abundancias} 
\end{center} 
\end{table*}

\subsubsection{Surface Abundances} 

In all of our calculations the internal solutions (of the Henyey
iteration) were calculated up to a fitting mass fraction
$\xi_1=$ln$(1-{m_r}/{M_\star})=-20$ during the proton burning and the
first years after it (with the exception of the sequence shown in
Fig. \ref{fig:BAtimestimeres} and \ref{fig:mu_ages}, where $\xi_1=-12$
).  This means that approximately $2 \times 10^{-9}\, M_\star$ are
left out of chemical integration. Even though this is enough to
account for the gravothermal energy generated in the outer layers of
the star, it may be still not small enough to follow in detail the
chemical evolution that takes place in the outermost surface of the
star. This prevents us from making a detailed comparison with the
observed abundances of Sakurai's object.  Indeed, convective dilution
as well as modest mass loss rates in the giant state are expected to
erode the last vestiges of H left in the star after the born-again
chemical processing, exposing the underlying H-deficient
layers. Nonetheless, we can still try to compare the abundances
characterizing the envelope of our models with those of the PG 1159
stars and V4334. The values of the surface abundances (i.e. the mass
fraction at the outermost shell of the interior) of our models is
presented in Table \ref{tab:abundancias} (for those calculated with a
MTS of $10^{-5}$ yr). Note that our surface abundances are
[He/C/O]=[0.38/0.36/0.22], in agreement with those obtained by Herwig
(1999). Also the abundances presented in Table \ref{tab:abundancias}
are similar to surface abundance patterns observed in PG1159 stars
([He/C/O]=0.33/0.50/0.17]) Dreizler \& Heber (1998), but with a lower
abundance of He and C, and a larger amount of oxygen.  These
differences in the abundances of He and O may be taken as a signal
that the overshoot parameter $f$ is a bit smaller than the one used
here. In fact, lower overshoot efficiency yields lower oxygen
abundances in the intershell region during the thermally pulsing AGB
phase (Herwig 2000). Also it is expected that intershell abundances
may depend on the mass of the remnant. It is worth noting the
remarkably good agreement between the $^{14}$N abundances of our
models and those observed in pulsating PG1159 stars (Dreizler \& Heber
1998). This makes these objects more likely to be produced by a VLTP,
in contrast with the situation for nonpulsators PG1159 which do not
show $^{14}$N features at their surface. As it is shown in Table
\ref{tab:abundancias} in all of the calculations (with different
numerical and physical parameters) we obtain stellar models that
display a surface $^{14}$N mass abundance of 0.01. This is exactly
what it is observed in PG1159 pulsators (and also on V4334 Sgr). This
reinforces the idea that PG1159 pulsators would have experienced a
VLTP in the past. On the other hand PG1159 nonpulsators could have
followed a different evolutionary channel, for instance a LTP (in
which hydrogen is only diluted as a consequence of the He flash, and
no $^{14}$N is produced). It is worth noting in view of the discussion
presented by Vauclair et al. (2005), that in no one of the sequences
we performed here an He-enriched surface composition with no
detectable nitrogen was obtained (typical composition of PG1159
nonpulsators). One may be tempted to think that if stellar winds were
stronger during the pre-PG1159 evolution models with no trace of
$^{14}$N could be obtained, but this mechanism would not lead to a
He-enriched surface.

In Fig. \ref{fig:surfacequimi} we show that our models predict a
second change in the surface abundance of H (the first happened during
violent proton burning), while the other elements remain almost
unchanged. This takes place when the star reaches the giant region and
a convective envelope starts to develop (see
Fig. \ref{fig:HR_full}). As a result of the non instantaneous mixing
at the (short lived) PBCZ, H abundance varies with depth (with higher
amounts of H at the top). Thus when the envelope starts to mix due to
convection the surface abundance of H is diluted and its value drops
by more than an order of magnitude. This is qualitatively similar to
the observed behavior of the surface abundances in Sakurai's object
during 1996 (Asplund et al. 1999), before it dissapeared as
consequence of dust episodes in 1998.

We note that lower values of $D$ increase the amount of He and lower
the surface oxygen. This is expected because mixing of material
becomes less eficient with a lower $D$. As was noted by Lawlor \&
MacDonald (2003), we also find that when the value of $D$ is reduced
by a factor of 100, the H abundance remaining at the surface of the
star is not consistent with the ones observed at the Sakurai's
object. Although, this inconsistency may be due to the fact that we are
not taking into account mass loss. Finally, we note that models with
small convective efficiency are characterized by surface $^{14}$N
abundances somewhat smaller than observed.

\section{Final remarks and conclusions}
In this paper we have studied the born again scenario for 0.5842 and
0.5885 \msun\ model star, under several numerical and physical
assumptions. For consistency purposes the VLTP initial stellar models
have been obtained following the complete evolution of initially 2.5
and 2.7 $M_\odot$ stellar models, from the ZAMS through the thermally
pulsing AGB phase. We incorporate an exponentially decaying diffusive
overshoot above and below any formally convective border during the
whole evolution. The inclusion of a time-dependent scheme for the
simultaneous treatment of nuclear burning and mixing processes due to
convection, salt finger and overshoot has allowed us a detailed study
of the abundance changes over all the evolutionary stages. Also we
incorporate in our study the double diffusive GNA convection theory
for fluids with composition gradients, which allows us to study
  the role of the chemical gradients.

 In the present work we have found that, according to our
treatment of mixing and burning, born again times are very sensitive
to the adopted time resolution during the proton ingestion, and that
these times converge to a given value when we keep the minimum allowed
time step (MTS) below $\sim 5\times10^{-5}$
 For the sequences
calculated with MTS $\lesssim 5\times10^{-5}$ we find that:

\begin{itemize}
\item The inclusion of $\nabla\mu$ in the calculation of the mixing
velocity during the violent proton burning leads to an increase in the
born again times by about a factor of 2. This leads to born again
times of 10-20 yr which are not consistent with the observed born
again time of Sakurai's Object.
\item One interesting point of the present work is that we find, with
the standard mixing length theory (i.e. not considering $\nabla\mu$ in
the calculation of mixing velocities), very short born again times of
5-10 years (that is of the order of magnitude of those observed in
V605 Aql and V4334 Sgr) even in the case that no reduction in the
mixing efficiency is invoked. Our born again times are then closer (although
shorter) to the ones originally found by Iben \& MacDonald (1995). 
\end{itemize}

 Also other important results of the present work are:
\begin{itemize}

\item We find that proton ingestion can be seriously altered if the
occurence of overshooting is modified by the $\nabla\mu$-barrier at
the H-He interface, and thus changing the born again times.
\item  Also we find that saltfinger
instability regions develop below the proton burning convective
zone. The occurrence of such instability regions bears no consequences
for the further evolution of the star, particularly the evolutionary
timescales.
\item A detailed description of the development of proton burning has
been provided. In particular we identify two different stages of the
proton ingestion.
\item We have compared the evolution of the surface parameters of the
models with those observed in Sakurai's Object and we find that,
although born again times are larger than the observed, both
luminosity curve and the surface cooling rate show a qualitative
agreement with those observed in Sakurai's Object. Also the drop in
the H abundance (with no significant change in the other chemical
species) observed during 1996 in Sakurai's object is qualitatively
predicted by our models. 
\end{itemize} 

 We close the paper by mentioning some of the work that need to
be done before we can finally understand the VLTP scenario. In
particular an assessment of the influence of the AGB progenitor
evolution (particularly regarding the number of thermal pulses
experienced by the progenitor star) on the born again times would be
valuable. Also calculations of VLTPs including a simultaneous
resolution of both structural and composition equations are needed.
But more importantly a more detailed treatment of convection than the
local and one dimensional approach attempted here is necessary. In this
connection two items should be studied. First a detailed study of the
interaction of chemical gradients and overshooting mixing is
in order. Second, as the timescale during H-ingestion is of the
order of magnitude of the convective turn over timescale, it would be
necessary to explore the H-flash on timescales at and below the
convective turn over timescale and to study convection away from the
diffusion limit.

\begin{acknowledgements}
We warmly acknowledge our referee F. Herwig for a careful reading of
 the manuscript. We are very thankful for his comments and suggestions
 which strongly improve the original version of this work. Part of
 this work has been supported by the Instituto de Astrofisica La
 Plata. A.M.S has been supported by the National Science Foundation through
 the grant PHY-0070928.

\end{acknowledgements}

\appendix

\section{Two Linearization Schemes for Stellar Structure Equations}

An appropriate choice of the  linearization scheme of the equations of stellar
structure  can be  of crucial  importance in  certain evolutionary  stages. In
order  to determine  the  importance  of such  a  choice in  the  case of  the
born-again scenario, two intrinsically different schemes were used. 

Following  the  notation of  Sugimoto  (1970),  we  write the  four  evolution
equations as 
\begin{equation}
{\partial y_i}/{\partial  x}=\phi(x,\vec{y}),
\end{equation}
where we denote by $y_i, \, i=1,2,3,4$ the dependent variables 
$\ln{p}$, $\ln{T}$,  $\ln{r}$ and $L$ (here  $p$, $T$, $r$ and  $L$ have their
usual meaning) and by $x$ the independent variable related to the fractional 
mass. 

The first aproach is the one proposed by Sugimoto (1970) for rapid evolutionary
phases, generally  characterized by negative temperature  gradients like those
developed during shell flashes below the point of maximum energy release. 
Here, the finite difference equations are written as:
\begin{equation}
{y_i}^{k+1}-{y_i}^k=\Delta  x^k   \left[\beta_i  \phi_i(y^k,x^k)+  (1-\beta_i)
  \phi_i(y^{k+1},x^{k+1})\right] 
\end{equation}
where $\beta_1=\beta_3=1/2$, $\beta_2=1$ and $\beta_4=0$. This means
that the derivatives of $\ln{T}$ and $L$ between the mesh points
$k+1$ and $k$ are approximated by their values at these points. As
shown by Sugimoto (1970) this representation  leads to a more
stable scheme during rapid evolutionary phases (as it is the case for the
born again  scenario). The gain in  stability is at  the expense of a  loss in
accuracy,  which   can  be  troublesome  when  long   evolutionary  steps  are
used.   However, as  shown by  Sugimoto (1970),  when the  time-step  is small
enough, the accuracy of the  usual implicit scheme ($\beta_i=1/2$
for  all $i$)  is recovered.  The  time-scale associated  with the  born-again
phase, ensure  that the  evolutionary time-step always  remain small,  and the
accuracy of the method is then granted.

The second linearization scheme is based on the representation of the 
finite-difference equations in two grids of mesh points in a de-centered
way. Specifically, on the first grid $x_j$ ($j=1,...,N$), we evaluate 
$L$ and $\ln{r}$, while $\ln{p}$ and $\ln{T}$ are specified on
a pseudo-grid ($j'=1,...,N$), defined by $x_{j'}=(x_j+x_{j-1})/2$
for $j'=2,...,N-1$ and by $x_{1'}= x_1$ and $x_{N'}=x_N$ (surface and center
respectively). Accordingly, the equations are written as:
\begin{eqnarray}
{y_i}^{k'}-{y_i}^{k'-1}= \Delta x^{k'}
\phi_i(\frac{{y_1}^{k'}+{y_1}^{k'-1}}{2}&,&\frac{{y_2}^{k'}+{y_2}^{k'-1}}{2}
\nonumber \\ 
, {y_3}^{k-1}&,&{y_4}^{k-1},x^{k-1}  ) 
\end{eqnarray}
for $i=1,2$, and
\begin{eqnarray}
{y_i}^k-{y_i}^{k-1} = \Delta x^k 
\phi_i({y_1}^{k'}&,&{y_2}^{k'},\frac{{y_3}^k+{y_3}^{k-1}}{2}\nonumber \\ 
&,&\frac{{y_4}^k+{y_4}^{k-1}}{2}, \frac{x^k+x^{k-1}}{2}) \nonumber 
\end{eqnarray}
for  $i=3,4$. This  method has  the advantage  that all  the  derivatives have
information from two mesh points while at  the same time it is very robust and
does  not give  rise to  unphysical solutions  as those  found with  the usual
implicit scheme in certain
evolutionary phases as mentioned above.  This is our preferred
choice for linearizing the stellar structure equations. 

In practice, born-again calculations were carried out using both schemes just
described.  The time-scales found with both schemes, while keeping the physical
inputs and other numerical issues (see Sect.~\ref{sec:numerics}) the same, 
were in all  our tests very similar, the differences  being smaller than 10\%,
and  the absolute  time-scale being  determined solely  by the  choice  of the
minimum time-step.

 \end{document}